\documentclass{article}
\usepackage{amssymb}
\usepackage{algorithmic}
\usepackage{graphicx}
\usepackage{hyperref}
\usepackage{multicol}
\usepackage{epstopdf}
\usepackage{placeins}
\usepackage{url}
\usepackage{alltt}
\usepackage{eurosym}
\usepackage{xfrac}
\usepackage{amsfonts}
\usepackage{bbm}
\usepackage{float}
\usepackage{color}
\usepackage{multirow}
\usepackage{amsmath}
\usepackage{amstext}
\usepackage{subfigure}
\usepackage{enumerate}
\usepackage[square,sort,comma,numbers]{natbib}
\usepackage{array}
\newcolumntype{P}[1]{>{\raggedright\arraybackslash}p{#1}}

\newcommand{\rp}[0]{R{\footnotesize APID}P{\footnotesize RO}M}
\newcommand{\prom}[0]{P{\footnotesize RO}M}

\AtBeginDocument{%
	\mathchardef\mathcomma\mathcode`\,
	\mathcode`\,="8000
}
{\catcode`,=\active
	\gdef,{\mathcomma\discretionary{}{}{}}
}
\mathchardef\breakingcomma\mathcode`\,
{\catcode`,=\active
	\gdef,{\breakingcomma\discretionary{}{}{}}
}

\begin{document}

%\bibliographyunit[\chapter]
%\defaultbibliographystyle{unsrt}

\providecommand\phantomsection{}

\newcommand{\rapidminer}[0]{R{\footnotesize APID}M{\footnotesize INER}}

\title{RapidProM: Mine Your Processes and Not Just Your Data}
\author{Wil M.P. van der Aalst, Alfredo Bolt, Sebastiaan J. van Zelst}

\maketitle

\begin{abstract}
The number of events recorded for operational processes is growing every year.
This applies to all domains: from health care and e-government to production and maintenance.
Event data are a valuable source of information for organizations that need to meet requirements related to 
compliance, efficiency, and customer service. 
Process mining helps to turn these data into real value:
by discovering the real processes,
by automatically identifying bottlenecks,  
by analyzing deviations and sources of non-compliance, 
by revealing the actual behavior of people, etc.
Process mining is very different from conventional data mining and machine learning techniques.
\prom\ is a powerful open-source process mining tool supporting hundreds of analysis techniques.
However, \prom\ does not support analysis based on scientific workflows.
\rp, an extension of \rapidminer\ based on \prom, combines the best of both worlds. 
Complex process mining workflows can be modeled and executed easily and subsequently reused for other data sets.
Moreover, using \rp, one can benefit from combinations of process mining 
with other types of analysis available through the \rapidminer\ marketplace. 
\end{abstract}

\section{Introduction}\label{sec:intro}
The number of events recorded in operational processes is growing every year.
This applies to all domains: from healthcare and e-government to production and maintenance.
Event data are a valuable source of information for organizations that need to meet requirements related to
compliance, efficiency, and customer service.
Within organizations the interest in data science and big data is rapidly growing.
Therefore they need to handle and analyze novel sources of data in smarter and more efficient ways.
However, in order to \textit{improve business processes and services} it is often not sufficient to focus on data storage and data analysis alone.
In fact, it is often the case that individual data elements within a process-based data set are not independent at all.
\emph{Process mining}~\cite{process-mining-book-2016} is a \textit{process centric} technique that helps to turn event data into real value:
by discovering the real processes,
by automatically identifying bottlenecks,
by analyzing deviations and sources of non-compliance,
by revealing the actual behavior of people, etc.

Process mining is a rapidly growing sub discipline within both Business Process Management (BPM)~\cite{ISRN-Spotlight-BPM-survey} and data science \cite{i-esa-keynote2014}.
Process mining is different from conventional data mining and machine learning techniques 
as it specifically takes into account that the event data originates from a business process.
Many mainstream data mining and machine learning techniques on the other hand neglect this fact, i.e., they are \emph{not} process centric.
As a result, mainstream Business Intelligence (BI), data mining and machine learning tools are not tailored towards the analysis of event data and the improvement of processes.
Fortunately, there are dedicated process mining tools able to transform event data into actionable process-related insights.
For example, \prom~\cite{2009_aalst_prom} (\url{www.promtools.org}) is an open-source process mining tool supporting analyses such as process discovery, conformance checking, social network analysis, organizational mining, clustering, decision mining, prediction, and recommendation.
Moreover, in recent years, several vendors released commercial process mining tools.
Examples include:
\emph{Celonis Process Mining} by Celonis GmbH	(\url{www.celonis.de}),
\emph{Disco} by Fluxicon	(\url{www.fluxicon.com}),
\emph{Interstage Business Process Manager Analytics} by Fujitsu Ltd	(\url{www.fujitsu.com}),
\emph{Minit}	by	Gradient ECM	(\url{www.minitlabs.com}),
\emph{myInvenio}	by	Cognitive Technology	(\url{www.my-invenio.com}),
\emph{Perceptive Process Mining}	by	Lexmark	(\url{www.lexmark.com}),
\emph{QPR ProcessAnalyzer}	by	QPR	(\url{www.qpr.com}),
\emph{Rialto Process}	by	Exeura	(\url{www.exeura.eu}),
\emph{SNP Business Process Analysis}	by	SNP Schneider-Neureither \& Partner AG	(\url{www.snp-bpa.com}), and
\emph{PPM	webMethods Process Performance Manager}	by	Software AG	(\url{www.softwareag.com}).
The growing number of process mining tools illustrates the relevance of process mining.

As mentioned above, \prom\ is a powerful open-source tool supporting hundreds of process mining analysis techniques.
However, \prom\ does not support the creation and execution of \textit{analytic workflows}.
We therefore recently introduced \rp~\cite{RapidProM-BPM2014-demo-CEUR}\footnote{\rp\ is an open source project, the source code is openly available via \url{http://github.com/rapidprom}} (\url{www.rapidprom.org}), a \rapidminer\ extension that wraps around the core functionality present within \prom.
It entails stable algorithms for the purpose of process mining analysis such as process discovery, conformance checking, performance analysis, etc.
\rp\ combines the advantages of the academic nature of \prom, i.e., it consists of state-of-the-art process mining algorithms, with the advanced data mining and analytic workflow capabilities of \rapidminer.
Complex process mining workflows can be modeled and executed easily and subsequently reused for other data sets.
Moreover, using \rp, one can benefit from combinations of process mining with other types of analysis available through the \rapidminer\ marketplace.

\begin{figure}[tb]
	\begin{center}
		\begin{center}
			\includegraphics[width=0.75\textwidth]{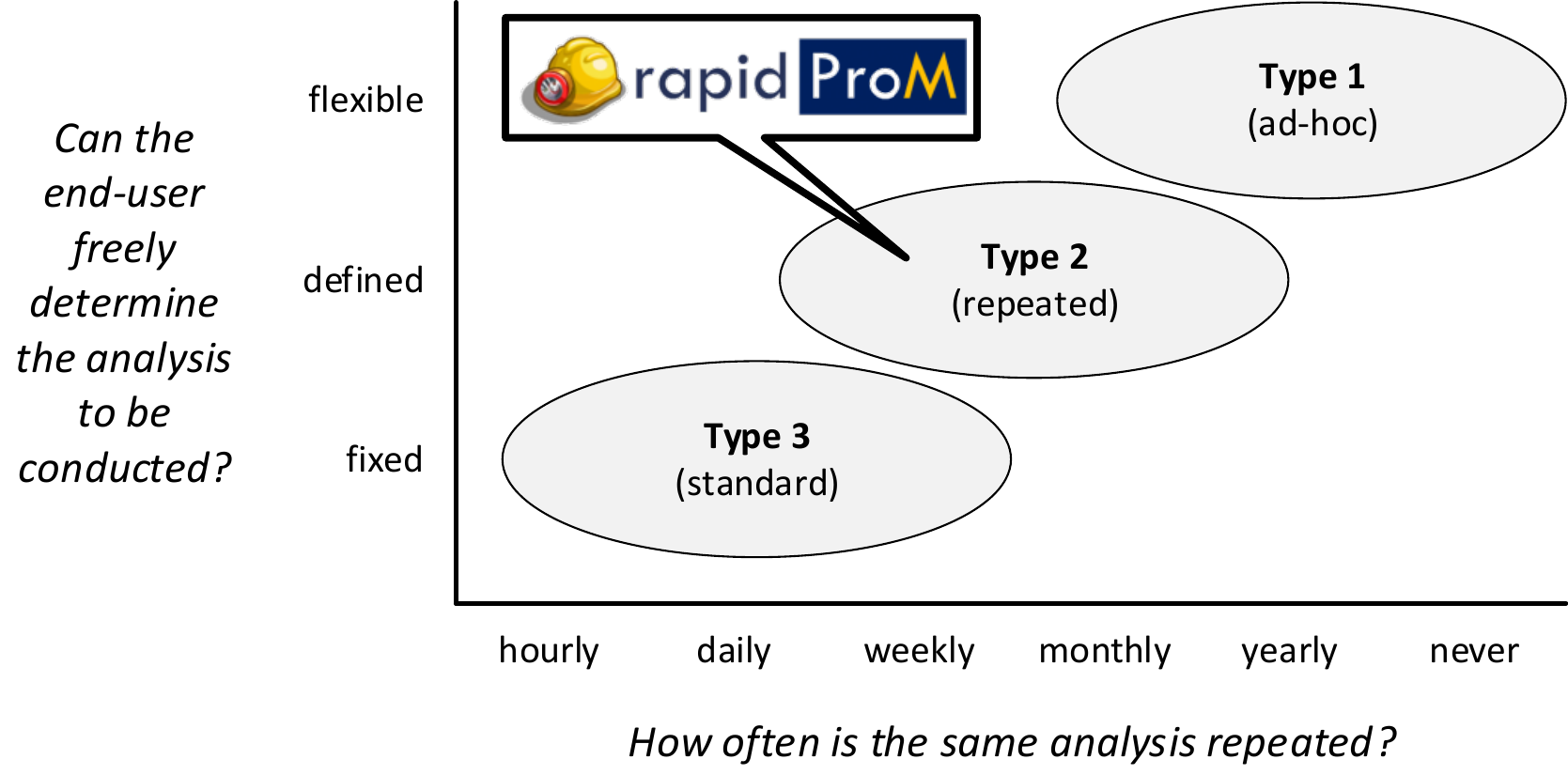}
		\end{center}
		\caption{Diagram showing three different categories of process mining tools and the positioning of \rp.}
		\label{fig:tool_types}
	\end{center}
\end{figure}

In~\cite{process-mining-book-2016}, three categories of process mining tools are identified, which are schematically depicted in \autoref{fig:tool_types}.
\textit{Type 1} process mining tools are mainly built for answering ad-hoc questions about business processes.
An example tool of such type is \textit{Disco}, which allows the user to interactively filter the data and project this immediately on a (newly learned) process model.
Tools of \textit{Type 3} are tailored towards answering predefined questions repeatedly in a known setting.
These tools are typically used to create ``process dashboards'' that provide standard views.
A tool like \textit{Celonis Process Mining} supports the creation of such process-centric dashboards.
Tools of \emph{Type 2} aim to answer questions that are recurring, but possibly at a lower frequency.
Analysis workflows may be predefined, but do not need to be completely fixed upfront.
Customization may be needed and the interpretation of the results requires knowledge
of process mining and understanding of the data. Unlike tools of \textit{Type 1} and \textit{Type 3},
the analytic workflow is made explicit such that it can be adapted and reused.
\rp\ is one of the few tools of \textit{Type 2}. 
It can also be used as a \textit{Type 1} tool, however, the workflow-like interaction style makes it predominantly a \textit{Type 2} tool.
\rp\ can also be used for process mining research where experiments need to be repeated for different data sets and parameter values.
Due to the \textit{loop} and \textit{subprocess} functionalities of \rapidminer,
\rp\ allows users to define a specific process mining analysis and repeat this multiple times
while varying parameters or event data.

The development of the first version of \rp\ started at the Eindhoven University of Technology around 2014 in context of the STW project ``Developing Tools for Understanding Healthcare Processes'' \cite{RapidProM-BPM2014-demo-CEUR}.
The extension consisted of basic operators for the purpose of process discovery, conformance checking, performance analysis, data exploration, etc.
Since then, \rp\ has been under active development, i.e., algorithms have been updated continuously, and, new operators were added.
In~\cite{rapid_prom_Alfredo_STTT}, \rp\ is used to assess the applicability of analytic workflows in a process mining context.
Additionally the work offers a set of basic workflow patterns for the purpose of process mining.

The goal of this chapter is to provide a basic overview of process mining, and, in particular \rp.
We assume that the reader is not familiar with process mining, but has used \rapidminer\ before.
The basic architecture is given and the most prominent objects and operators are discussed.
We also present three case studies showing both the usefulness and applicability of applying process mining by using analytic workflows.

The remainder of this chapter is organized as follows.
In \autoref{sec:procmin} we present a basic overview of the field of process mining.
In \autoref{sec:procdiff} we briefly argue the main differences between conventional data mining and process mining.
In \autoref{sec:rp} we present \rp\ by means of an overview of its architecture, common objects and operators.
In \autoref{sec:casestudies} we present three case studies highlighting different functionalities of \rp.
Finally, \autoref{sec:concl} concludes the chapter.

\section{What is Process Mining?}\label{sec:procmin}
The main goal of process mining is to improve operational processes by using event data.
By exploiting recorded event data, process mining techniques are able to show \textit{what actually} happened during the execution of a business process.
We identify three main types of process mining, being \textit{process discovery}, \textit{conformance checking} and \textit{performance analysis}.
We present each of these types in more detail.
Prior to this, we introduce the main source of data used within process mining, i.e., \textit{event logs}.
Throughout this section we use a simplified data set based on a real ``Road Traffic Fine Management'' process originating from an Italian municipality.
%We introduce the event log, accompanied by some simplified analyses to explain the main components of process mining.
In \autoref{sec:casestudies} we go more in depth and perform three case studies based on the real data set using \rp.

\subsection{Event Logs}
Process mining is impossible without proper \emph{event logs} \cite{process-mining-book-2016}.
Fortunately, event logs can be extracted from a wide variety of data sources, including enterprise systems (SAP, Oracle, etc.),
hospital information systems (Epic, ChipSoft, Siemens, etc.),
middleware (IBM, Microsoft, etc.), business process management systems,
mobile applications, social media, sensor networks, etc.
Most databases contain event information. However, often quite some efforts are needed to extract event logs from them (scoping, selection, and transformation).  

An event log contains event data related to a particular process.
Each \emph{event} in an event log refers to an activity executed for a particular \emph{process instance}, also referred to as a \emph{case}.
Events related to a case are ordered and can have any number of additional attributes.
Examples of typical attributes next to the mandatory case identifier, activity and time attributes are resource, costs, transaction type, location, etc.
Not all events need to have the same set of attributes.
However, typically, events referring to the same activity have the same set of attributes.
\begin{figure}[h]
	\begin{center}
		\includegraphics[width=1\textwidth]{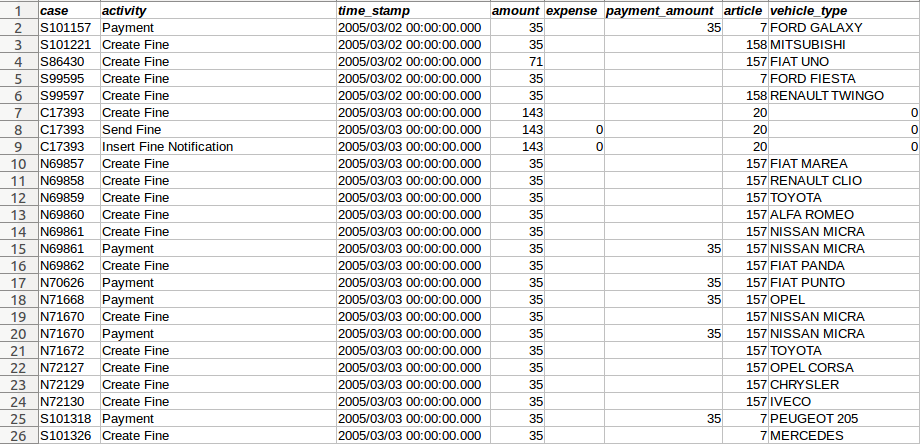}
	\end{center}
	\caption{Simplified event log where each line refers to an \textit{event} in the process of handling road fines.}
	\label{fig:event_log}
\end{figure}

A simplified example of an event log, based on a real event log, is depicted in \autoref{fig:event_log}.
The event log shows a few events related to handling traffic fines.
Consider for example \textit{row 2} in the example, i.e., the first event below the header.
This event relates to a \textit{Payment} activity, preformed in context of case \textit{S101157}.
The \textit{amount} of the fine equals the \textit{amount of payment}, as indicated by the \texttt{amount} and \texttt{payment\_amount} columns respectively.
Note that additional information such as the \textit{law article} related to the fine as well as the \textit{vehicle type} is also recorded.
\autoref{fig:event_log} shows three events that are all related to case \textit{C17393} (\textit{rows 7,8,9}).
Three activities are performed in sequence for this case, i.e., \textit{Create~Fine}, \textit{Send~Fine}, and, \textit{Insert~Fine~Notification}.
Together these three events form a \textit{trace} describing the lifecycle of a traffic fine.

\begin{figure}[tbh]
	\begin{center}
		\includegraphics[width=1\textwidth]{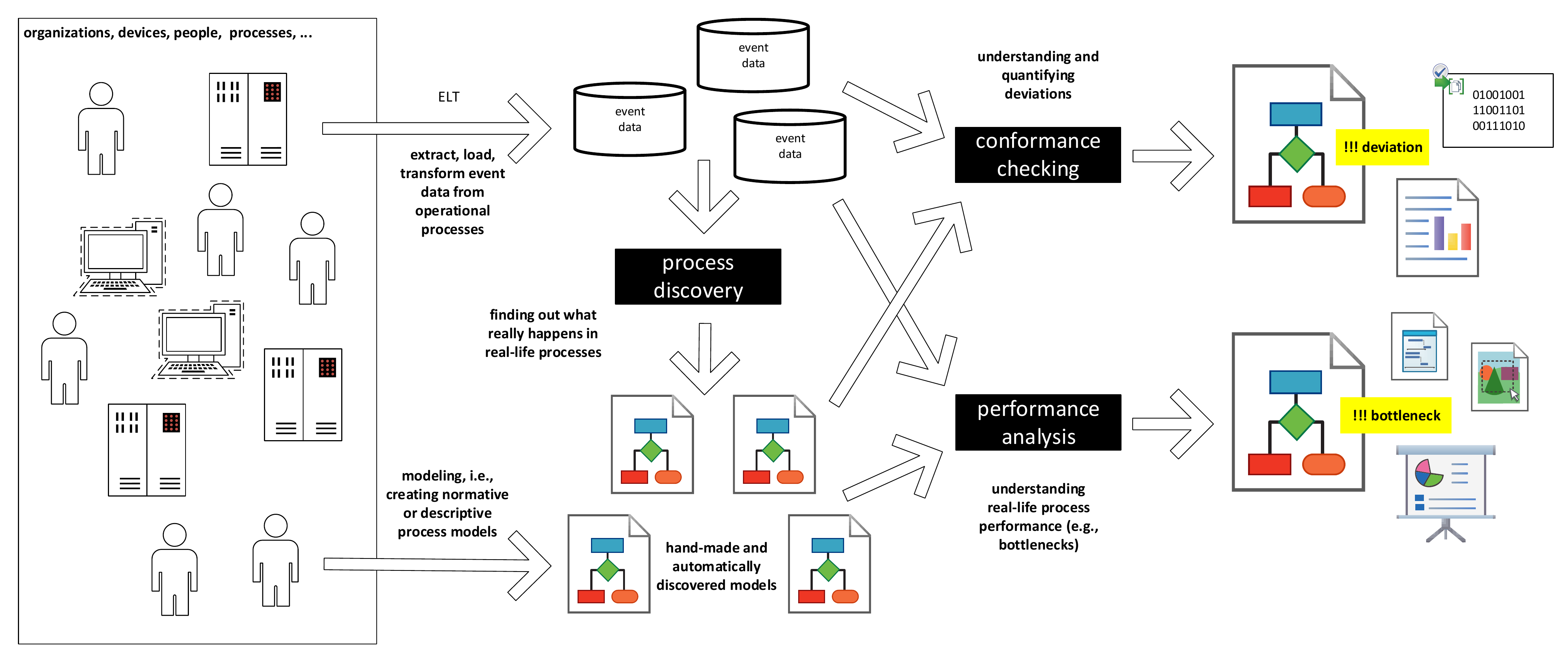}
		\caption{Schematic overview of process mining spectrum showing the three main types of process mining: 
			(1) process discovery, (2) conformance checking, and (3) performance analysis.}\label{fig-procmin}
	\end{center}
\end{figure}
\autoref{fig-procmin} shows an overview of process mining highlighting the three main types of analysis.
Event logs are represented by the three shapes labeled with \textit{event data}.
In the upcoming sections we discuss each sub discipline, i.e., process discovery, conformance checking, and, performance analysis in more detail.
As the figure illustrates, all these techniques use, amongst other artifacts, event logs as an input.

\subsection{Process Discovery}
\label{sec:procmin:processdiscovery}
The first type of process mining is discovery (see \autoref{fig-procmin}).
A process discovery technique takes an event log as an input and produces a model without using any a-priori information.
Typically the focus of process discovery techniques in on the \textit{control-flow} aspect of a process.
Within this aspect we are mainly interested in what ways the activities within the process can be ordered.
Several process model formalisms exist, e.g., BPMN~\cite{BPMN-OMG-2}, Petri nets~\cite{murata} etc.
\begin{figure}[tbh]
	\begin{center}
		\includegraphics[width=1\textwidth]{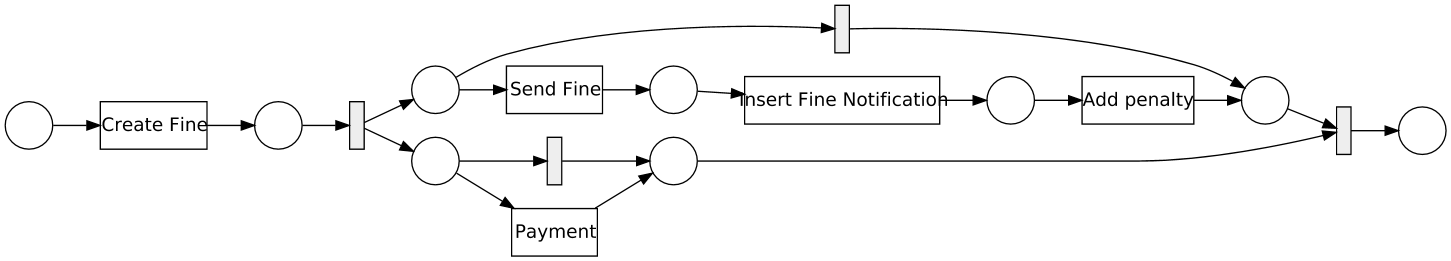}
		\caption{Simplified Petri net describing the main behavior present in the example event log of \autoref{fig:event_log}.}
		\label{fig:example_process_model}
	\end{center}
\end{figure}

A simplified example of a process model, using the Petri net formalism, is depicted in \autoref{fig:example_process_model}.
This process model was discovered based on the event log depicted in \autoref{fig:event_log}.
Within the model, a square with an inscription represents an activity.
The squares without inscription refer to \textit{invisible activities}.
Such invisible activities are usually used for routing purposes, or, to allow us to skip certain activities, i.e., we are able to skip the \textit{Payment} activity.
This particular process model describes that within the process, a fine is always created first.
After this two \textit{parallel branches of behavior} are started.
The upper branch describes that the sequence of activities \textit{Send~Fine}, \textit{Insert~Fine~Notification} and \textit{Add~penalty} should be performed, or, the activities are not performed at all.
The lower branch specifies that a payment should be performed, or it should be skipped.
If all activities within the two branches are performed or skipped, the process finishes.
Note that the fact that the two branches are executed in parallel implies that the execution of 
the payment activity is independent of the execution of the activities in the upper branch.
Thus, according to this process model, the payment activity can be performed at any point in time, though after the fine is created.

\subsection{Conformance Checking}
\label{sec:procmin:conformancechecking}

The second type of process mining is conformance checking (see \autoref{fig-procmin}).
Here, an existing process model is compared with an event log of the process that the model is describing, i.e.,
modeled behavior is confronted with observed behavior.
Conformance checking can be used to check if reality, as recorded in the log, conforms to the model and vice versa.
\begin{figure}[tbh]
	\begin{center}
		\includegraphics[width=1\textwidth]{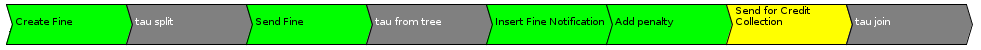}
		\caption{An example conformance checking diagnostics for a particular process instance.}
		\label{fig:example_conformance}
	\end{center}
\end{figure}

Consider again the simple process model depicted in \autoref{fig:example_process_model}.
As an example, we take the following actual trace from the event log used as a basis of the snapshot in \autoref{fig:event_log}: 
$\langle \textit{Create~Fine}, \textit{Send~Fine}, \textit{Insert~Fine~Notification}, \textit{Add~penalty}, \textit{Send~for~Credit~Collection} \rangle$.
Obviously, this trace is not completely conforming with respect to the model as it contains activity \textit{Send~for~Credit~Collection}, which is not present in the model.
Applying conformance checking techniques for this trace results in \autoref{fig:example_conformance}.\footnote{In \autoref{fig:example_conformance} we project conformance diagnostics onto a trace. However, conformance diagnostics can also be projected on the process model.}
The green blocks indicate that the algorithm was able to match an event in the trace to an activity in the model.
The gray blocks correspond to the invisible activities as introduced in the previous section.
Finally the yellow block corresponds to a non-conforming event.
In this case, it refers to an activity that did occur in the trace of events, though the model does not describe the activity.
Such type of non-conforming event is often referred to as a \textit{log move}.
Clearly, it is also possible that the model describes that a certain activity should happen, though the activity is not present in the trace of events.
In such case we refer to a \textit{model move}.

\subsection{Performance Analysis}
\label{sec:procmin:performanceanalysis}
The third type of process mining is performance analysis by replaying the event data on a discovered process model (see \autoref{fig-procmin}).
Typically, techniques from conformance checking are used combined with timing information present in the event log, e.g., by using timestamps we are able to deduct the average time in-between two events etc.
These types of statistics are subsequently projected onto a discovered or hand-made process model.
\begin{figure}[tbh]
	\begin{center}
		\includegraphics[width=1\textwidth]{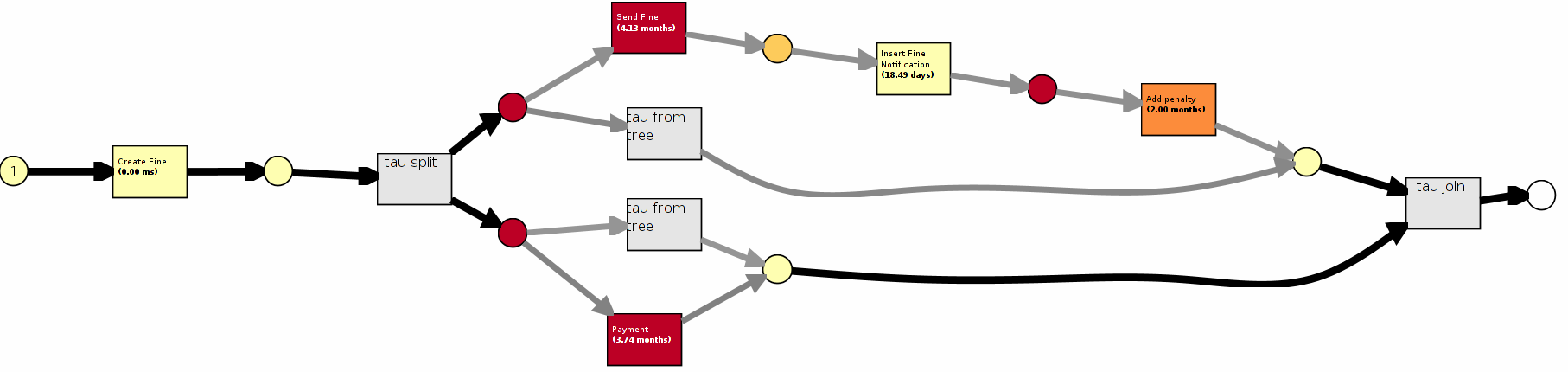}
		\caption{By replaying the event log on the process model, it is possible to annotate the model with performance information. 
			The red transitions and places correspond to bottlenecks in the process.}
		\label{fig:example_performance_analysis}
	\end{center}
\end{figure}

The results of performance analysis can be used to identify the problematic parts of the process, e.g., bottlenecks.
An example of such analysis is depicted in \autoref{fig:example_performance_analysis}.
Within the figure, we projected timing information from our example log on the process model of \autoref{fig:example_process_model}.
In this example, the yellow colored elements of the process indicate efficient flows, whereas orange and/or red elements highlight activities that are taking longer.
Again, the gray components correspond to invisible activities and are not considered within the performance analysis.

\subsection{Other Types of Process Mining}
\begin{figure}[tbh]
	\begin{center}
		\includegraphics[width=0.5\textwidth]{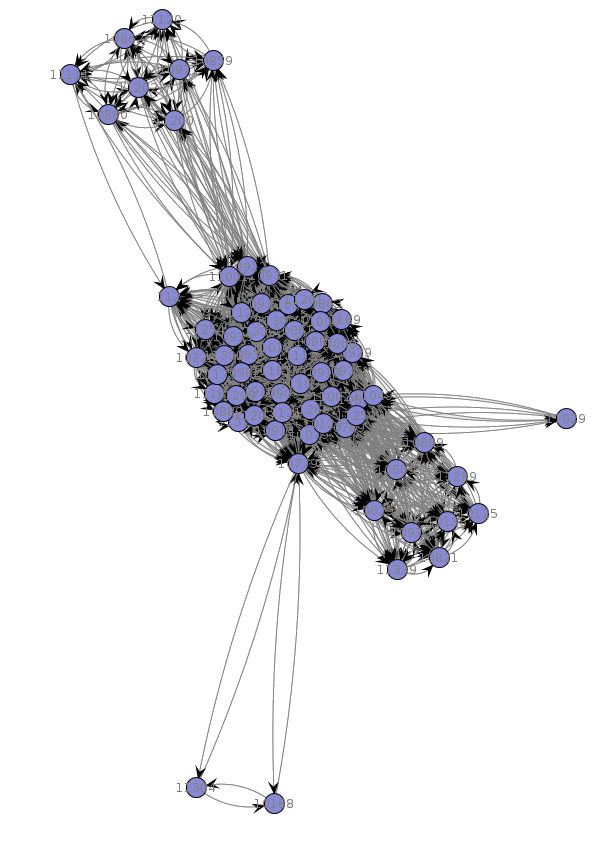}
		\caption{An example of a social network based on similarity of work profiles of resources.}
		\label{fig:social_network}
	\end{center}
\end{figure}

Apart from the three subfields mentioned in the previous sections, other analyses have been developed within the domain of process mining as well.
An example of analysis is predictive analysis, i.e., predicting the remaining time of a process instance.
Also, some analysis methods focus on the organizational perspective of the process.
Usually an event log contains information regarding what resource executed what activity.
Hence, several interesting social networks can be mined from the event log.
Consider for example \autoref{fig:social_network} which depicts a \textit{similar task social network}.

The vertices within the social network correspond to resources that where active within the process.
An arc between two vertices indicates that the resources execute similar tasks within the process.
Clearly there are a few clusters identifiable within the network, suggesting that different groups of 
resources execute a different set of activities. 

Another type of process mining is decision mining which focuses on the data flow in a process model.
For each decision point, one can derive explanations based on different features. 
This can be used to learn that fines of a particular type, e.g., fines for speeding versus fines related to parking illegally, correlate with specific process executions.

\section{Why is Process Mining Different From Data Mining?}\label{sec:procdiff}

Process mining provides a range of tools to improve processes in a variety of
application domains. In the previous section we introduced the main forms of process mining.
Process mining builds on process model-driven approaches and data mining. However, process
mining is much more than an amalgamation of existing approaches. For example,
existing data mining techniques are too data-centric to provide 
a comprehensive understanding of the end-to-end processes in an organization.
Some would argue that process mining is part of the broader data mining or machine learning discipline.
Depending on the definition, this could be (partially) correct.
A common definition for data mining is ``the analysis of (often large) data sets to find unsuspected
relationships and to summarize the data in novel ways that are both
understandable and useful to the data owner'' \cite{hand-mannila-smyth2001}.
Using this broad definition, parts of process mining are indeed included.
However, discussions on such inclusion relations are seldom useful and are
often politically motivated \cite{process-mining-book-2016}. 
Most data mining tools do \emph{not} provide process mining
capabilities, most data mining books do \emph{not} describe process mining techniques,
and it seems that process mining techniques like conformance checking do \emph{not} fit
in any of the common definitions of data mining. It is comparable to claiming that
``data mining is part of statistics''. Taking the transitive closure of both statements,
we would even be able to conclude that process mining is part of statistics. Obviously,
this does not make any sense.

Like process mining, data mining is data-driven.
However, unlike process mining, mainstream data mining techniques are typically
not process-centric. Process models expressed in terms of Petri nets or BPMN diagrams
cannot be discovered or analyzed in any way by the main data mining tools.
Typically, data mining techniques assume that the data items used originate 
from some unknown distribution, and moreover, are \textit{independent}.
Within process mining, the primary source of data is an event log.
At the lowest level, an event log consists of events, however, multiple events together constitute to a trace.
Taking into account that events together constitute to cases effectively adds an additional layer 
within the data which can not be ignored.
Thus, within process mining we do not only consider the events in isolation, rather, 
we also look at events at the trace level and use this to gain new insights within the data.

There are a few data mining techniques that come close to process mining \cite{process-mining-book-2016}. Examples
are sequence and episode mining \cite{sequence-mining-agrawal-95,mannila97}. 
However, these techniques do not consider end-to-end processes.

One of the important features of \rp\ is that the standard data mining techniques shipped with \rapidminer\ can be combined 
with a range of process-centric analytical techniques from \prom.
Through process mining, it becomes easier to apply data mining techniques to event data. 
The process model serves as a backbone for a variety of data mining techniques (classification, clustering, etc.).
For example, decision rules can be learned using standard
data mining tools after the control-flow backbone (e.g., a Petri net) has been
learned using a process mining tool.
It is also interesting to combine process mining with other types of analysis available though the \rapidminer\ marketplace 
(e.g., text mining, web mining, deep learning, etc.).

\section{RapidProM}\label{sec:rp}
In this section we describe \rp, the process mining extension of \rapidminer.
We present a high-level view of the basic architecture of the extension, commonly used objects, and, an overview of commonly used operators.

\subsection{Architecture}\label{sec:rp-arch}
As indicated in the introduction, the process mining toolkit \prom~\cite{2009_aalst_prom,process-mining-book-2016} is the standard scientific tool for performing process mining analytics.
As a consequence, \prom\ consists of a vast amount of state-of-the-art algorithms for the purpose of process discovery, conformance checking and performance analysis.
Conveniently, both \prom\ and \rapidminer\ are programmed in the \texttt{java} (\url{http://www.java.com}) programming language.
Hence, porting existing algorithms implemented in \prom\ is mainly concerned with integrating them within the \rapidminer\ ecosystem.

\begin{figure}[!t]
	\begin{center}
		\includegraphics[width=0.65\textwidth]{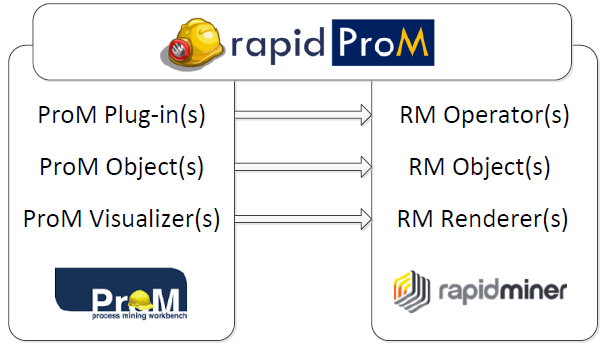}
		\caption{Basic overview of the architecture of \rp.}\label{fig-arch}
	\end{center}
\end{figure}

In \autoref{fig-arch} a high-level overview of the architecture of \rp\ is depicted schematically.
\rp\ acts as a bridge between \rapidminer\ and \prom.
Within \prom\ any algorithm that performs some process mining task, e.g., a process discovery algorithm is called a \textit{plug-in}.
Such plug-in usually expects a (set of) input object(s) and results in a (set of) output object(s).
Thus in that respect a plug-in is comparable with an \textit{operator} as defined within \rapidminer, and, most operators in \rp\ correspond directly to their counterpart in \prom.
Each plug-in has an associated (set of) visualizer(s) that allow the user to inspect the results of their process mining analysis.
Again, most of these visualizers are ported as \textit{renders} into \rp\, and hence, have a corresponding counterpart in \prom.

\subsection{Objects}\label{sec:rp-obj}
Objects specific to \rp\ are either based on data stored in some source, e.g., files, or, are the result of applying some algorithm.
A complete overview of all objects ported and/or defined in \rp\ is outside the scope of this chapter.
However, we present the most prominent categories of objects relevant for process mining related analyses.

\textbf{Event Logs.}
The most prominent data objects within \rp\ are event logs.
As indicated, an event log is a collection of traces which represent process instances that have been executed.
Several operators within \rp\ need an event log as an input object.
\rp\ supports importing data files which adhere to the \texttt{XES} standard (\url{http://www.xes-standard.org/}).
Most commonly \texttt{.xes} files are used, an \texttt{XML} based data source adhering to the standard.
Within \rp, the OpenXES reference implementation is used for importing/exporting \texttt{.xes} files.
Additionally, \rp\ supports converting \texttt{example sets} into event log objects, and, event log objects into \texttt{example sets}.\footnote{An \texttt{example set} is the basic tabular data object in \rapidminer.}

\textbf{\textit{Process Model Related Objects.}}
Since most algorithms within process mining are related to (business) processes, several objects implementing some process model formalism are available in \rp.
Examples of such formalisms are BPMN Models~\cite{BPMN-OMG-2-formal}, Petri nets~\cite{murata}, colored Petri nets~\cite{jensen_2009_cpn}, process trees~\cite{sander-scalable-BPMDS2015}, etc.

\textbf{\textit{Analysis and Reporting Objects.}}
Apart from event logs and process model related objects, \rp\ includes several objects related to analysis and reporting.
For example, we are able to perform conformance checking of event logs and Petri nets by computing so called \textit{alignments}~\cite{wires-replay}.
Hence, \rp\ includes \prom\ based alignment objects with corresponding visualizations to inspect conformance results.
Another example would be the \textit{Inductive Visual Miner}~\cite{Inductive_Visual_Miner-BPM2014-demo-CEUR} object, which presents a visual animation of cases going through a process model. This object is interactive, as it enables the user to filter cases based on selected behavior. It also highlights deviations of cases w.r.t. the process model.

\subsection{Operators}\label{sec:op}
\rp\ introduces several operators that allow us to perform process mining analyses.
We present the most prominent categories of process mining operators.

\textbf{\textit{Input/Output.}}
For some of the objects defined in \rp\ \textit{import} and \textit{export} operators are available, e.g., Event logs, Petri nets, etc.
All these operators need the input/output file/folder as a parameter.
Additionally, for event logs an \textit{extractor} operator is present which is able take a file as an input object.

\textbf{\textit{Discovery.}}
\rp\ provides several different process discovery algorithms, e.g., the Alpha Miner~\cite{aal_min_TKDE}, the Heuristics Miner~\cite{tonbeta166}, and the Inductive Miner~\cite{sander-tree-disc-PN2013}.
Apart from these \textit{control-flow} based algorithms, i.e. algorithms focusing on the sequential ordering of business activities within a process, \rp\ also provides a Social Network Miner~\cite{aal_sna_cscw}.
This operator allows us to discover a social network of interacting resources within the event log.
The corresponding renderer allows the user to inspect and manipulate this network in an interactive fashion.

\textbf{\textit{Analysis.}}
\rp\ provides several operators for the purpose of analyzing process discovery results.
Several conformance checking related operators are present that allow the user to assess the conformance of an event log to a process model and vice versa.
Furthermore, several operators that enable the user to analyze/enhance process models are provided, 
e.g., for analyzing formal properties of the models and/or simplifying the models.
Finally, \rp\ offers operators that allow the user to visualize actual data present in the event log onto the model, e.g., ``replaying'' the event log on a process model.

Apart from the three main categories mentioned here there are some more operators available in \rp.
There are operators that provide the possibility to convert certain objects into other objects, e.g., Petri nets to BPMN models.
Other operators allow the user to manipulate event logs, e.g., adding artificial start and end events to traces within event logs (which greatly improves the performance of some process discovery algorithms).

\begin{figure}[tbh]
	\begin{center}
		\includegraphics[width=1\textwidth]{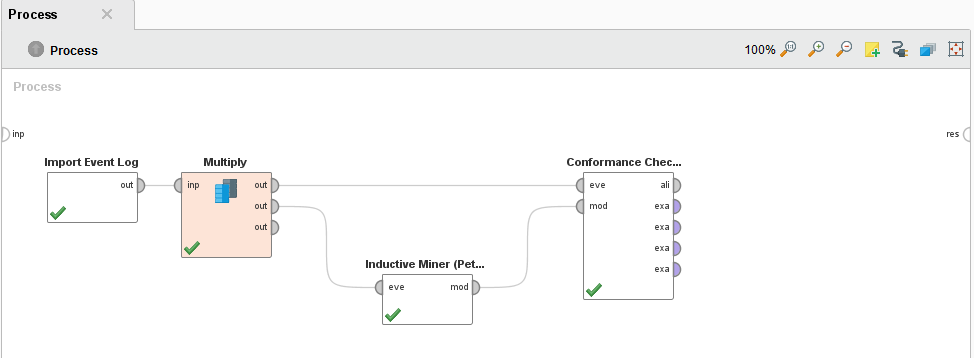}
		\caption{An example of a \rp\ workflow where (i). an event log is imported, (ii). a process model is discovered based on the event log (using the Inductive Miner), and, (iii). a conformance checking analysis is performed using both the event log and the resulting process model.}
		\label{fig:example_discovery_conformance}
	\end{center}
\end{figure}

An example workflow containing an import operator, a discovery operator and an analysis operator is depicted in \autoref{fig:example_discovery_conformance}.

\section{RapidProM in Action}\label{sec:casestudies}
This section describes three concrete use cases in which the operators provided by \rp\ are used to obtain insights about a process using the process mining techniques described in \autoref{sec:procmin}.
We aim at using simple workflows in order to show the main \textit{process mining capabilities} of \rp.
A real-life event log, containing more than 150.000 cases, is used as input for all three use cases.
The event log is obtained from the data archive of the 4TU.Centre for Research Data (4TU.ResearchData).\footnote{\url{http://data.4tu.nl}}
Within this archive, other event logs are publicly available as well, i.e., \url{https://data.4tu.nl/repository/collection:event_logs} as part of the ``IEEE Task Force on Process Mining - Event Logs'' collection.
Within the case studies we do not go into detail w.r.t. the algorithmic properties of the operators used.
However, the complete documentation (i.e., operator description, parameter explanation and related articles) of all the operators used within the workflows described in the case studies is provided by \rp\ and is embedded in \rapidminer\ (by means of the Help Window).
Finally, note that some of the results depicted in this section are enhanced to increase their readability on paper and therefore in some cases might deviate from their corresponding visualization(s) in \rp.

\subsection{Dataset: Road Fines Management Process}
\label{sec:cs1d}
The dataset used in the three case studies is an event log that was extracted from an information system that handles the ``Road Traffic Fine Management'' process in an Italian municipality.\footnote{This event log is publicly and permanently available at (\url{http://dx.doi.org/10.4121/uuid:270fd440-1057-4fb9-89a9-b699b47990f5}).}
The event log includes  cases of road traffic fines that are processed by the municipality over a period of three years (from January 2000 until June 2013).
The ``Road Traffic Fine Management'' process is composed of 11 different activities.
We describe the process as it \textit{should} look like according to domain experts. 
The activities, as recorded within the event log, are written italic.

\begin{figure}[tbh]
	\begin{center}
		\includegraphics[width=0.44\textwidth]{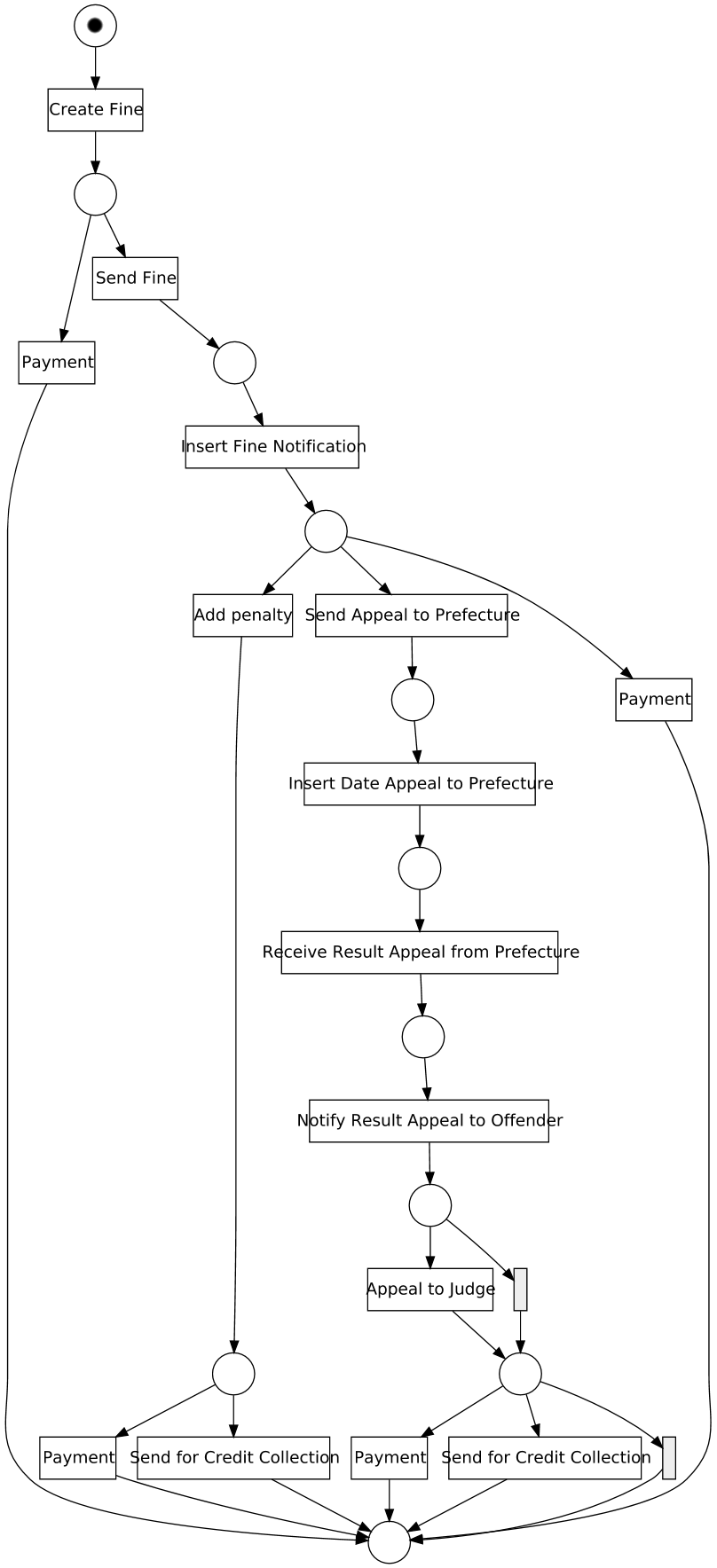}
	\end{center}
	\caption{Normative process model based on the ``Road Traffic Fine Managment'' process description.}
	\label{fig:case_study_normative}
\end{figure}

The process starts with a fine being created (i.e., \textit{Create Fine}).
After a fine has been created, it is sent to the offender's place of residence (i.e., \textit{Send Fine}).
When the offender receives the fine, the date of reception of such notification is also registered (i.e., \textit{Insert Fine Notification}).
After this, the fine should be paid within 60 days (i.e., \textit{Payment}).
However, in case the fine was physically handed over to the offender, e.g. by means of a parking ticket, the offender is able to immediately pay the fine.
In such case, the fine will not be sent and there will be no registration of the notification.
This exception, i.e., direct payment after ticket creation saves the offender administration costs.
In total, the offender has 60 days to either pay the fine or appeal against it.
After this period, a penalty is added to the fine amount (i.e., \textit{Add Penalty}).
If the offender appealed against the fine within 60 days, the appeal is sent to the corresponding prefecture (i.e., \textit{Send Appeal to Prefecture}), which is registered when it is received (i.e., \textit{Insert Date Appeal to Prefecture)}.
The results of the appeal are sent back to the municipality (i.e., \textit{Receive Result Appeal from Prefecture}) and they are notified to the offender (i.e., \textit{Notify Result Appeal to Offender}), which can appeal against the result (\textit{Appeal to Judge}).
If the offender does not pay (possibly after a denied appeal), the fine is sent for credit collection (i.e., \textit{Send for Credit Collection}).

A simple normative process model (Petri net) based on the description in the previous paragraph is depicted in \autoref{fig:case_study_normative}.
Note that this model does not capture any form of parallelism.
It is merely a sequence of activities combined with possible choices, e.g., after the \textit{Create Fine} activity, the process model describes the choice of either a \textit{Payment} activity, or, a \textit{Send Fine} activity.

\subsection{Case Study 1: Discovering Process Models}\label{sec:cs1}

As mentioned in \autoref{sec:procmin:processdiscovery}, many different models can be discovered from the same event log by applying different discovery techniques.
Furthermore, process models can be described using different notations, e.g., using Petri net, BPMN, etc.
This case study shows how several process models can be discovered from a real event log using the \textit{discovery} operators of \rp.

\begin{figure}[t]
	\begin{center}
		\includegraphics[width=0.6\textwidth]{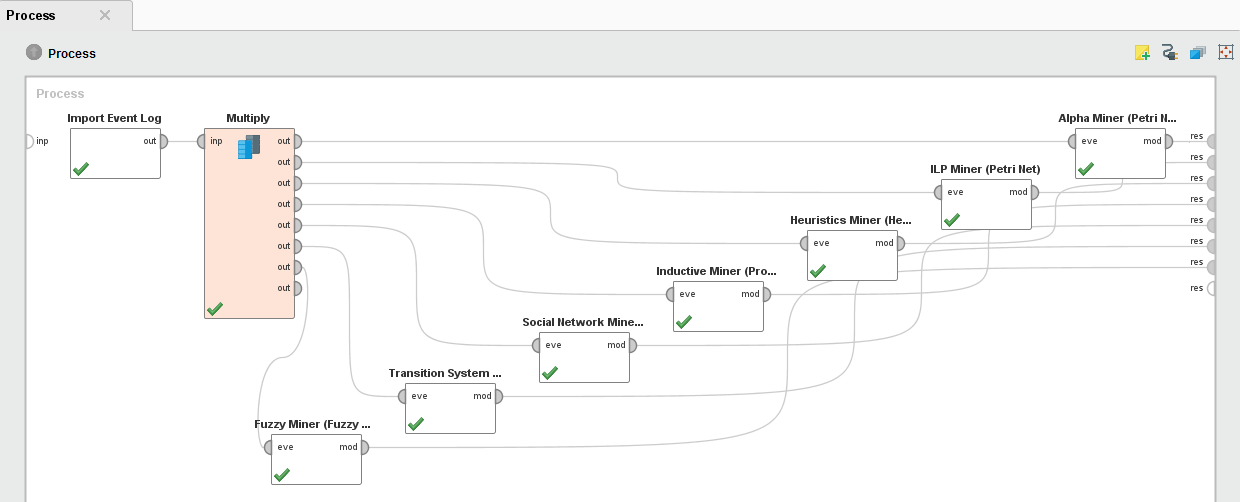}
		\caption{Workflow used for Case Study 1: an event log is imported and several process model are created using some of the available discovery operators provided by \rp.}
		\label{fig:caseStudy1Workflow}
	\end{center}
\end{figure}

\textbf{\textit{Workflow}}
\autoref{fig:caseStudy1Workflow} illustrates a basic \rapidminer\ workflow used to discover process models from an event log.
First, an event log is imported using the \textit{Import Event Log} operator.
Then, several process models are discovered from the event log using different operators that
produce process models in different notations.
The Discovery operators included in this workflow, and the type of process model that they produce are:
the \textit{Alpha Miner} (Petri net),
the \textit{ILP Miner} (Petri net),
the \textit{Heuristics Miner} (heuristics net),
the \textit{Inductive Miner} (Petri net),
the \textit{Social Network Miner} (social network),
the \textit{Transition System Miner} (transition system), and
the \textit{Fuzzy Miner} (fuzzy model).

The workflow immediately highlights an advantage of using \rapidminer\ as it allows us to generate all results of the different miners in one go.
Moreover we are able to directly reuse this workflow for other event logs as well.
In such case we just change the file pointed to by the \textit{Import Event Log} operator.
If we do this, depending on the event log, we might need to change the \textit{classifier} used by the discovery algorithms.
The classifier parameter is shared by all discovery operators.
It is a default attribute within an event log and specifies what \textit{event attribute} should be used in order to identify the corresponding \textit{activity}.
In case of our example event log snapshot, depicted in \autoref{fig:event_log}, the classifier to use is the \texttt{activity} column.\footnote{Classifiers originate from the XES standard. More information on classifiers can be found in: \url{http://www.xes-standard.org/_media/xes/xesstandarddefinition-2.0.pdf}.}
Finally, the workflow can also be used as a sub-process after which the best model, depending on some process model specific quality criteria, e.g., model simplicity, is selected for further analysis.

\begin{figure}[b]
	\includegraphics[width=1\textwidth]{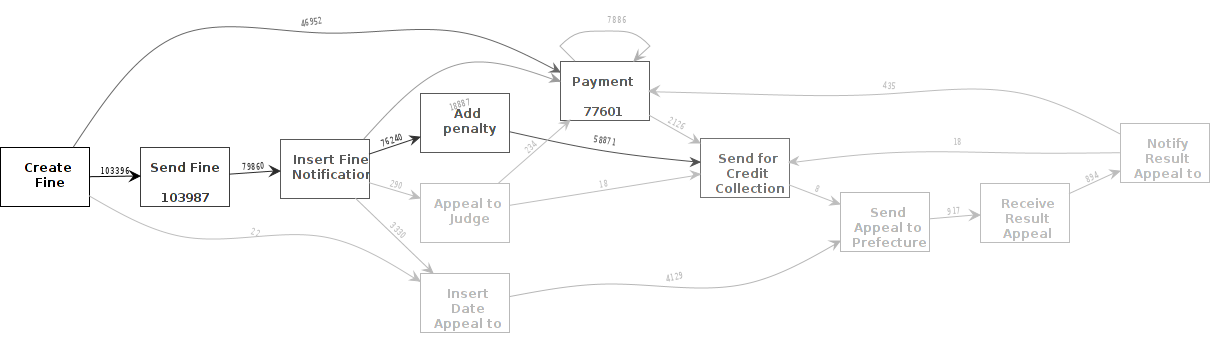}
	\caption{Result (heuristics net) of applying the Heuristics Miner on the event log.}
	\label{fig:case_study_1_heuristics}
\end{figure}

\textbf{\textit{Results Analysis}}
Consider \autoref{fig:case_study_1_heuristics}, in which the result of applying the Heuristics Miner on the event log is depicted.
The model depicted in \autoref{fig:case_study_1_heuristics} is a Heuristics net.
The squares refer to activities whereas an arrow between two activities indicates that the source activity preceded the target activity.
The intensity of the different elements within the net correspond to their relative frequency as observed within the event log.
Although the presence of parallelism is somewhat hard to detect in a heuristic net, it still provides very usable insights.
From the heuristics net we deduce that indeed, as prescribed in the textual and normative models, after the \textit{Create Fine} activity a \textit{Payment} activity can be performed.
However, when considering the \textit{Payment} activity, we observe a \textit{self-loop}, i.e. an arc leaving and entering the \textit{Payment} activity.
This possibly suggests that the specific municipality allows for paying the fine in terms.
Another interesting observation is the fact that after a penalty is added, the activity \textit{Send for Credit Collection} always seems to occur.
In some cases, even though a payment was made, still the \textit{Send for Credit Collection} activity was performed.
This can have several explanations, e.g., the payment did not match the fine, the payment was registered too late etc.
Finally after the \textit{Insert Fine Notification} activity, the \textit{Appeal to judge} activity seems to be executed.

\begin{figure}[t]
	\includegraphics[width=1\textwidth]{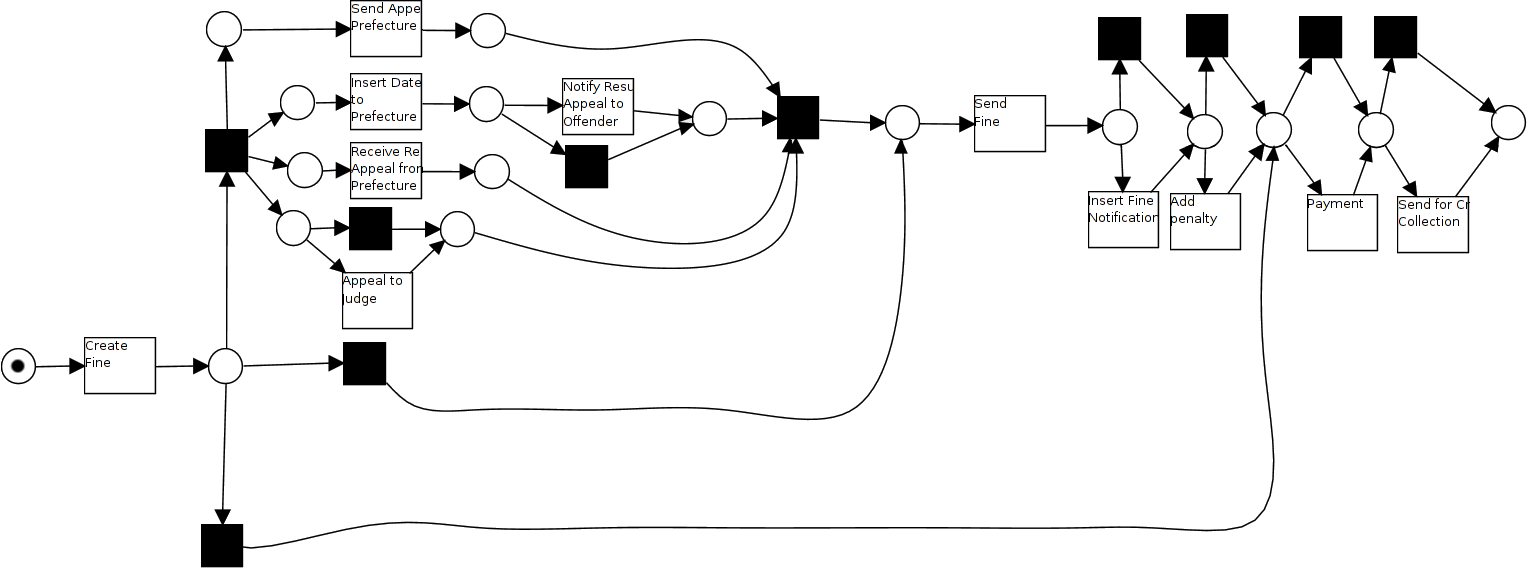}
	\caption{Result (Petri net) of applying the Inductive Miner on the event log.}
	\label{fig:case_study_1_inductive}
\end{figure}

Consider the Petri net depicted in \autoref{fig:case_study_1_inductive}, obtained by applying the Inductive Miner on th event log (using a noise filter level of 0.2).
Due to the application of a noise filter, the Petri net might neglect some of the behavior as captured within the event log.
The model describes that after the \textit{Create Fine} activity we are able to choose between executing three invisible activities (the black squares, intended for routing purposes).
Choosing the lowest invisible activity leads us directly to the \textit{Payment} activity.
We are however able to skip the activity.
If we execute the middle invisible activity (after the Create Fine activity), we always have to execute the \textit{Send Fine} activity.
After this the \textit{Insert Fine Notification} activity is either executed or skipped.
Finally, if we choose the upper invisible transition, the model shows some interesting behavior.
The activities \textit{Send Appeal to Prefecture}, \textit{Insert Date to Prefecture} and \textit{Receive Result Appeal from Prefecture} should all be executed.
However, the model describes that they are in a parallel block, i.e., we are able to executed them in any arbitrary order.
Note that the model also describes that we are able to only execute the \textit{Create Fine} activity and then skipping all subsequent activities, i.e., resulting in a trace with only one activity.

\subsection{Case Study 2: Checking Process Conformance}\label{sec:cs2}
The previous case study shows how process discovery can be used to get an insight in a process based on the \textit{actual behavior} recorded an event log.
Very often, as was the case in the previous case study, the event log contains behavior that does not comply with the process description.
This may have several reasons, e.g., the process description is incomplete, errors occurred during logging the process, the process was executed incorrect etc.
At the same time, due to limitations of the process discovery algorithm, a resulting process model may describe behavior that is \textit{not present in the event log}. 
Although applying process discovery yields insights w.r.t. the degree of compliance, it does not provide an accurate exact result.
Therefore, in this case study, we perform conformance checking of the normative process model w.r.t. the event log to get a more exact conformance result.

\begin{figure}[t]
	\begin{center}
		\includegraphics[width=0.6\textwidth]{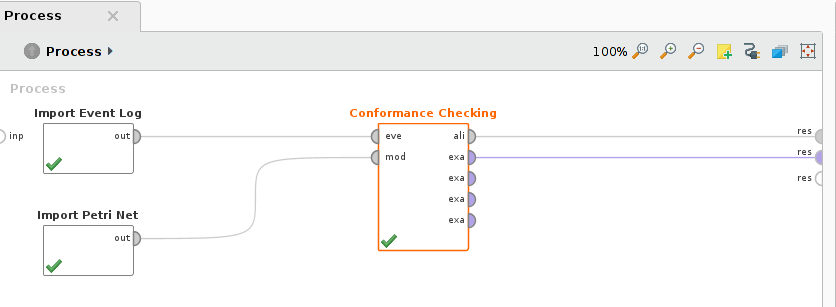}
		\caption{Workflow used for Case Study 2: an event log and Petri net are imported. Subsequently the model and the event log are used for conformance checking.}
		\label{fig:caseStudy2Workflow}
	\end{center}
\end{figure}

\textbf{\textit{Workflow}}
\autoref{fig:caseStudy2Workflow} illustrates the basic workflow for the purpose of conformance checking.
We import an event log using the \textit{Import Event Log} operator.
Secondly we import the normative model using the \textit{Import Petri net} operator. 
The two artifacts are used to check conformance by means of the \textit{Conformance Checking} operator.
Note that the \textit{Conformance Checking} operator needs a Petri net and an event log as an input, hence, the Petri net can also be the result of any process discovery operator (given that the operator produces a Petri net).
Hence, we can use the result of a conformance checking operator as a quality measure for a process discovery algorithm.
The \textit{Conformance Checking} operator provides several results, e.g., a projection of conformance results onto the process model, or, onto the event log. 
Also, conformance checking results are provided as \texttt{example sets} for further analysis using other \rapidminer\ operators.

%As we mentioned before, process mining results are specially difficult to illustrate in a book. This time, the reason is because most of these results contain interactive elements whose behavior cannot be flattened to a static image.
%In order to truly interact with each result, we encourage the reader to download and run the workflow in \rapidminer.

\begin{figure}[tbh]
	\centering
	\includegraphics[width=1\textwidth]{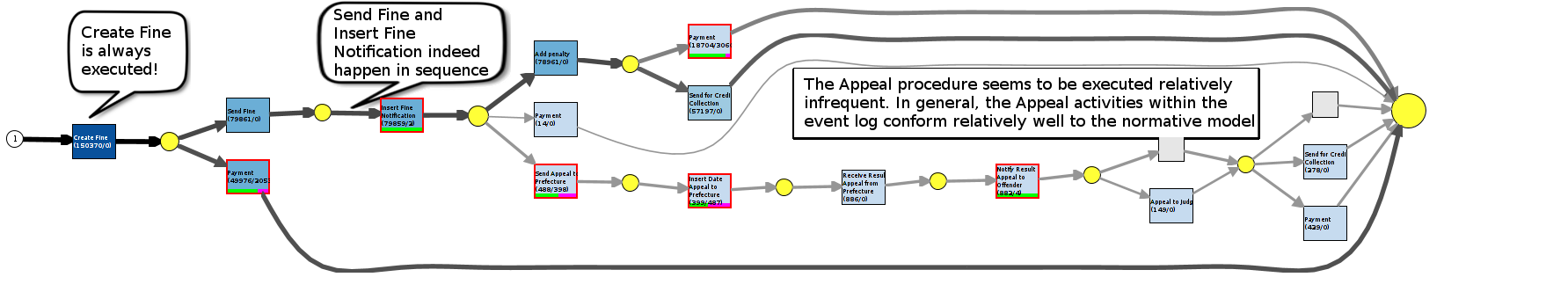}
	\caption{Conformance results projected onto the normative model}
	\label{fig:case_study_2_conf_model}
\end{figure}

\begin{figure}[b]
	\centering
	\begin{tabular}[t]{cc}
		\subfigure[Activity for which the event log and model conform perfrectly.]{\includegraphics[width=0.35\textwidth]{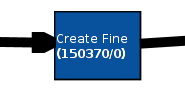} \label{fig:case_study_2_synchronous}}
		&
		\subfigure[Activity for which the event log and model \textit{do not} conform perfrectly.]{\includegraphics[width=0.35\textwidth]{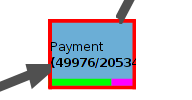} \label{fig:case_study_2_non_synchronous}}
	\end{tabular}
	\caption{Two activities from the normative model having a different type of conformance result.}
	\label{fig:case_study_2_types_transitions}
\end{figure}

\textbf{\textit{Results Analysis}}
In this section we analyze some of the results obtained by the \textit{Conformance Checker} in the workflow described in the previous section.
We first inspect the projection of the conformance checking results onto the process model, as depicted in \autoref{fig:case_study_2_conf_model}.
Note that the model depicted in \autoref{fig:case_study_2_conf_model} is the same as the normative model depicted in \autoref{fig:case_study_normative}, the difference in layout is related to the underlying visualization software.
Within \autoref{fig:case_study_2_conf_model} we identify two different types of squares, which we highlight in \autoref{fig:case_study_2_types_transitions}.

The first type of activity are highlighted in~\ref{fig:case_study_2_types_transitions}a, are highlighted in a solid blue color.
For these type of activities, the event log and the model are perfectly conforming w.r.t. each other.
The intensity of the color is related to the frequency of occurrence of such an activity.
As highlighted in \autoref{fig:case_study_2_conf_model}, the \textit{Create Fine} activity of the normative model aligns perfectly w.r.t. the event log (and vice versa).
This mainly implies that the activity is executed in each trace.
It does however not necessarily imply that the \textit{Create Fine} activity is indeed always the first activity within any sequence.
The exact reason for this is somewhat technical and outside of the scope of this chapter, hence, we refer the interested reader to~\cite{wires-replay} for more detail.
The \textit{Create Fine} activity is also depicted in \ref{fig:case_study_2_types_transitions}a.
The $(150370/0)$ inscription underneath the \textit{Create Fine} label indicates that in $150.370$ cases the event log and the model agreed that the activity should be executed, whereas in $0$ cases this was not the case.
Another example of a perfectly conforming activity is the \textit{Send Fine} activity.

The second type of activity are activities that are not perfectly conforming.
These activities have a red border, and, a green and pink bar in the bottom of the square.
Consider \autoref{fig:case_study_2_types_transitions}b for an example.
The width of the green bar indicates the number of conforming occurrences of the particular activity.
The pink bar indicates the number of non conforming occurrences.
In case of the \textit{Payment} activity, depicted in \autoref{fig:case_study_2_types_transitions}b, we observe that in $49.976$ cases the event log and model conform w.r.t. executing the activity.
However, in $20.534$ cases, the model dictates that the activity should occur, whereas according to the event log, this is not the case.

\begin{figure}[htb]
	\centering
	\includegraphics[width=0.8\textwidth]{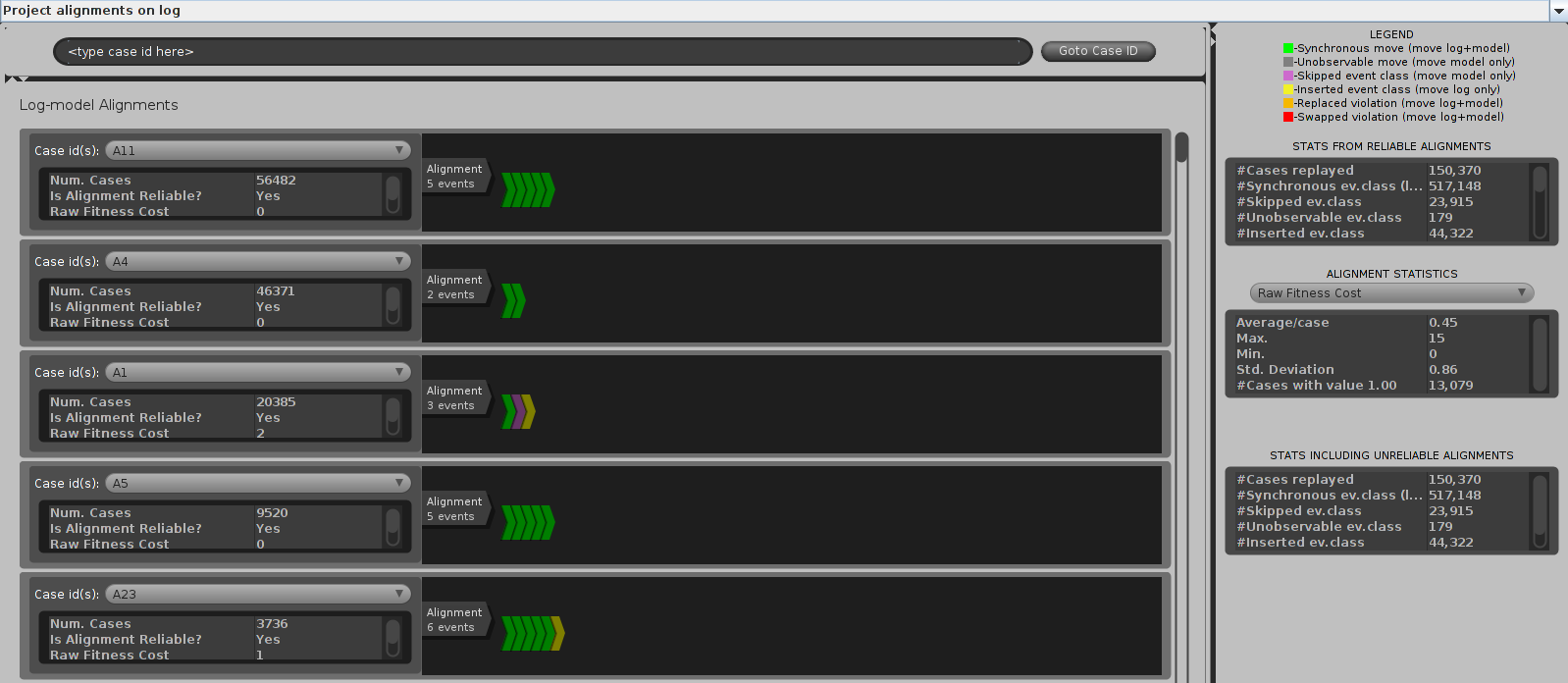}
	\caption{Example screen shot of conformance statistics projected onto cases in the event log.}
	\label{fig:case_study_2_example_conf_trace}
\end{figure}

\begin{figure}[b]
	\centering
	\includegraphics[width=1\textwidth]{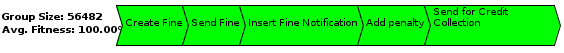}
	\caption{Conformance results projected onto a trace which perfectly conforming w.r.t. normative model.}
	\label{fig:case_study_2_perfect_trace}
\end{figure}

Although the projection of conformance metrics onto the model leads to interesting insights, it does not allows us to inspect the conformance of individual traces within the event log.
Moreover, as argued before, we are not able to deduce that the \textit{Create Fine} activity is indeed always the first activity of any case.
Within \rp, we therefore also provide the option to project the conformance results onto case within the event log (by means of a different \textit{renderer} of the conformance results).
An example screen shot of this view in \rp is depicted in \autoref{fig:case_study_2_example_conf_trace}.
Within this view, the cases are ordered based on their frequency within the event log.
Consider \autoref{fig:case_study_2_perfect_trace} depicting the first case of \autoref{fig:case_study_2_example_conf_trace} in more detail.
As the statistics show, there is a total of 56.482 traces within the event log that exactly follow this execution path.
Clearly this is a typical case of non-paying offenders that did not appeal.

\begin{figure}[t]
	\centering
	\includegraphics[width=1\textwidth]{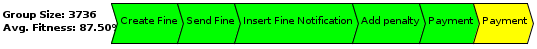}
	\caption{Conformance results projected onto a trace which does not conform w.r.t. the normative model.}
	\label{fig:case_study_2_non_fitting_trace}
\end{figure}

There are also cases for which the trace and the model are not conforming w.r.t. each other.
Consider the example depicted in \autoref{fig:case_study_2_non_fitting_trace}.
The figure indicates that indeed, the \textit{Create Fine} activity is both present in the trace and in the model.
Secondly, it indicates that a \textit{Send Fine} activity was performed, followed by an \textit{Insert Fine Notification} activity.
Somehow, the offender did not pay in time and hence a penalty was added to the fine.
Up until this point the trace and the model still conform.
However, after the first \textit{Payment} activity, the offender again performed a \textit{Payment} activity.
The second \textit{Payment} is not in line with the normative model and is therefore indicated in yellow.
There may be several reasons for a duplicate payment, e.g., the offender payed the ticket too late and afterwards payed the fine as well, the offender payed in multiple terms etc.
In any case, given that this behavior occurred around 3.736 times, such case is interesting to discuss with the business owner of the process, the Italian municipality in this case.

\subsection{Case Study 3: Identifying Bottlenecks in a Process}\label{sec:cs3}

The previous case study shows how to do conformance checking using an event log and a process model. 
In this case study, we go one step further and analyze the behavior in the event log from a \textit{performance} perspective, as described in \autoref{sec:procmin:performanceanalysis}.

\textbf{\textit{Workflow}}
The workflow used within this case study equals the workflow used within the previous case study.
However, instead of using the \textit{Conformance Checking} operator, we use the \textit{Analyze Performance (Manifest)} operator.
The result of this operator is a projection of performance statistics onto the input model, in this case the normative model.

\begin{figure}[b]
	\begin{center}
		\includegraphics[width=0.5\textwidth]{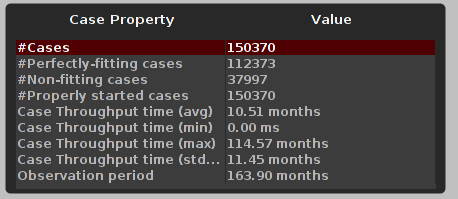}
		\caption{Screens shot of the global performance statistics of the normative process as reported by \rp.}
		\label{fig:case_study_3_global_statistics}
	\end{center}
\end{figure}

\textbf{\textit{Results Analysis}}
Some global statistics related to the performance of the process are depicted in \autoref{fig:case_study_3_global_statistics}
The average throughput time reported is 10.51 months, with a standard deviation of around 11.45 months.
Interestingly the longest running case was running for 114.57 months.
The reason for this can be a long running appeal against the fine. 
However, often these type of cases are related to issues in data quality, e.g. inaccurate logging of events etc. 
Thus, this stresses the need for proper data cleaning and filtering methods.

\begin{figure}[t]
	\begin{center}
		\includegraphics[width=1\textwidth]{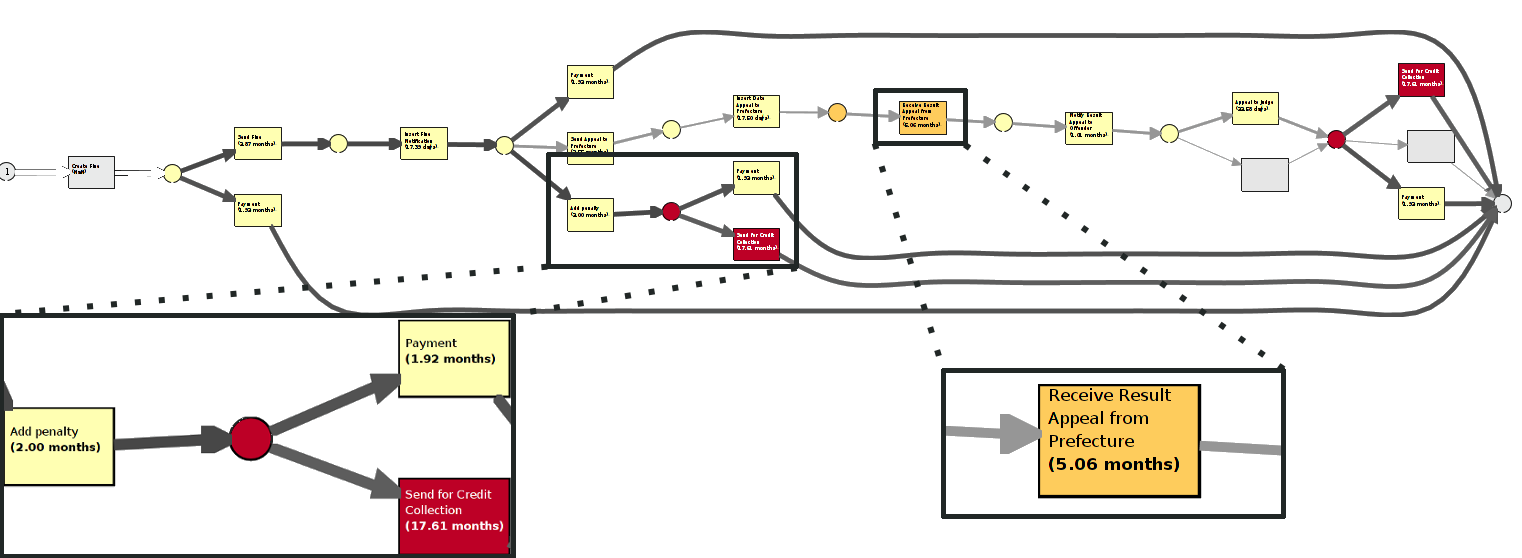}
		\caption{Projection of performance information into the normative process model.}
		\label{fig:caseStudy3Result1}
	\end{center}
\end{figure}

To gain more insights into the performance of individual activities we show the result of projecting performance information onto the normative process model in \autoref{fig:caseStudy3Result1}.
The timing information in the event log is projected onto the activities present in the model.
The activities in the process model are colored depending on their performance, i.e., red colors indicate longer waiting times and yellow colors indicate shorter waiting times. 

The result object is highly interactive: users can filter cases based on their throughput time, choose a different performance annotationm e.g., sojourn time, waiting time, change color schemes, among many other functionalities.
Also, for each activity several statistics are available such as minimum time, maximum time, mean, standard deviation etc.

The activity with the highest waiting time is \textit{Send for Credit Collection}, having a waiting time of 17.61 months on average. 
This high waiting time can be explained by the fact that only those fines that are not paid are sent to credit collection. 
Also, the time that the municipality waits before sending an unpaid fine for credit collection depends on the actions of the offender (e.g., if he/she appealed against the fine).

The second highest waiting time corresponds to the activity \textit{Receive Result Appeal from Prefecture}, with 5.06 months on average.
It seems that a large amount of time is spent between applying for an appeal and receiving the result from Prefecture.
Whether the municipality is able to speed up this part of the process is questionable, since this is most probably executed by an external party, i.e., the prefecture.

An interesting observation is the fact that activity \textit{Add Penalty} has a waiting time of 2.0 months on average.
This is in line with the fact that the municipality adds a penalty 60 days after the fine was sent. 
The standard deviation for this particular activity is around 30 minutes.
Due to this small deviation it is highly likely that adding the penalty is an \textit{automated activity}.
Another interesting observation is related to payments.
The average waiting time is 1.92 months, just under 2 months, i.e., the regular term for paying a fine.
Although this seems a reasonable period, the standard deviation is 4.08 months, indicating that the average wait time for payments is not strictly related to the legal payment term.

As the previous insights highlight, performing performance analysis using \rp\ leads to interesting results.
These results can subsequently be used to select and filter the event log in order to inspect the bottleneck cases in more detail.

\section{Conclusion}\label{sec:concl}
In this chapter we introduced the \rp\ plugin which extends \rapidminer\ with process mining capabilities.
\rp\ offers support to organizations that need to manage non-trivial operational processes.
Most of the operators provided by \rapidminer\ correspond to data handling and classical forms of 
data mining. These operators are \emph{not} process-centric and cannot be used to analyze and improve end-to-end processes. 
The operators in \rp\ provide \textit{process centric} analysis capabilities.
The \rp\ operators focus on the analysis of event data and process models.
They take into account that events are related to process instances and should be handled as such.
\rp\ supports discovering process models from event logs, checking conformance of an event log w.r.t.\ a process model, 
and calculating performance results based on an event log and a process model.
Combined with the powerful capabilities of \rapidminer\ in terms of building analytical workflows and data mining, \rp\ enables us to:
\begin{enumerate}
	\item Design complex process mining experiments, combining several different techniques.
	\item Reuse earlier defined analyses to be applied to new event logs (of the same or another process).
	\item Bridge the gap between process mining and data mining.
\end{enumerate}

We aim at keeping \rp\ up-to-date as new process mining techniques or more efficient implementations emerge.
For user/developer documentation, data sets, example workflows, etc., we refer to \url{www.rapidprom.org}.

\section*{Acknowledgements}
The authors would like to thank all that contributed the \prom\ plug-ins used in \rp\ (see \url{www.promtools.org}).
Special thanks go to Ronny Mans who developed the first version of \rp.

\bibliographystyle{unsrt}
%\bibliography{bibliography}

\begin{thebibliography}{10}
	
	\bibitem{process-mining-book-2016}
	{W.M.P. van der} Aalst.
	\newblock {\em {Process Mining: Data Science in Action}}.
	\newblock Springer-Verlag, Berlin, 2016.
	
	\bibitem{ISRN-Spotlight-BPM-survey}
	{W.M.P. van der} Aalst.
	\newblock {Business Process Management: A Comprehensive Survey}.
	\newblock {\em ISRN Software Engineering}, pages 1--37, 2013.
	\newblock doi:10.1155/2013/507984.
	
	\bibitem{i-esa-keynote2014}
	{W.M.P. van der} Aalst.
	\newblock {Data Scientist: The Engineer of the Future}.
	\newblock In K.~Mertins, F.~Benaben, R.~Poler, and J.~Bourrieres, editors, {\em
		Proceedings of the I-ESA Conference}, volume~7 of {\em {Enterprise
			Interoperability}}, pages 13--28. Springer-Verlag, Berlin, 2014.
	
	\bibitem{2009_aalst_prom}
	{W.M.P. van der} Aalst, {B.F. van} Dongen, C.W. G{\"{u}}nther, A.~Rozinat,
	H.M.W. Verbeek, and T.~Weijters.
	\newblock {ProM: The Process Mining Toolkit}.
	\newblock In {\em Proceedings of the Business Process Management Demonstration
		Track (BPMDemos 2009)}, 2009.
	
	\bibitem{RapidProM-BPM2014-demo-CEUR}
	R.~Mans, {W.M.P. van der} Aalst, and E.~Verbeek.
	\newblock {Supporting Process Mining Workflows with RapidProM}.
	\newblock In L.~Limonad and B.~Weber, editors, {\em Business Process Management
		Demo Sessions (BPMD 2014)}, volume 1295 of {\em CEUR Workshop Proceedings},
	pages 56--60. CEUR-WS.org, 2014.
	
	\bibitem{rapid_prom_Alfredo_STTT}
	A.~Bolt, {M. de} Leoni, and {W.M.P. van der} Aalst.
	\newblock {Scientific Workflows for Process Mining: Building Blocks, Scenarios,
		and Implementation}.
	\newblock {\em International Journal on Software Tools for Technology
		Transfer}, pages 1--22, 2016.
	
	\bibitem{BPMN-OMG-2}
	OMG.
	\newblock {Business Process Model and Notation (BPMN)}.
	\newblock Object Management Group, dtc/2010-06-05, 2010.
	
	\bibitem{murata}
	T.~Murata.
	\newblock {Petri Nets: Properties, Analysis and Applications}.
	\newblock {\em Proceedings of the IEEE}, 77(4):541--580, April 1989.
	
	\bibitem{hand-mannila-smyth2001}
	D.~Hand, H.~Mannila, and P.~Smyth.
	\newblock {\em {Principles of Data Mining}}.
	\newblock MIT press, Cambridge, MA, 2001.
	
	\bibitem{sequence-mining-agrawal-95}
	R.~Agrawal and R.~Srikant.
	\newblock {Mining Sequential Patterns}.
	\newblock In {\em {Proceedings of the 11th International Conference on Data
			Engineering (ICDE'95)}}, pages 3--14. IEEE Computer Society, 1995.
	
	\bibitem{mannila97}
	H.~Mannila, H.~Toivonen, and A.I. Verkamo.
	\newblock {Discovery of Frequent Episodes in Event Sequences}.
	\newblock {\em Data Mining and Knowledge Discovery}, 1(3):259--289, 1997.
	
	\bibitem{BPMN-OMG-2-formal}
	OMG.
	\newblock {Business Process Model and Notation (BPMN)}.
	\newblock Object Management Group, formal/2011-01-03, 2011.
	
	\bibitem{jensen_2009_cpn}
	K.~Jensen and L.M. Kristensen.
	\newblock {\em {Coloured Petri Nets - Modelling and Validation of Concurrent
			Systems}}.
	\newblock Springer, 2009.
	
	\bibitem{sander-scalable-BPMDS2015}
	S.J.J. Leemans, D.~Fahland, and {W.M.P. van der} Aalst.
	\newblock {Scalable Process Discovery with Guarantees}.
	\newblock In K.~Gaaloul, R.~Schmidt, S.~Nurcan, S.~Guerreiro, and Q.~Ma,
	editors, {\em Enterprise, Business-Process and Information Systems Modeling
		(BPMDS 2015)}, volume 214 of {\em Lecture Notes in Business Information
		Processing}, pages 85--101. Springer-Verlag, Berlin, 2015.
	
	\bibitem{wires-replay}
	{W.M.P. van der} Aalst, A.~Adriansyah, and {B. van} Dongen.
	\newblock {Replaying History on Process Models for Conformance Checking and
		Performance Analysis}.
	\newblock {\em WIREs Data Mining and Knowledge Discovery}, 2(2):182--192, 2012.
	
	\bibitem{Inductive_Visual_Miner-BPM2014-demo-CEUR}
	S.J.J. Leemans, D.~Fahland, and {W.M.P. van der} Aalst.
	\newblock {Process and Deviation Exploration with Inductive Visual Miner}.
	\newblock In L.~Limonad and B.~Weber, editors, {\em Business Process Management
		Demo Sessions (BPMD 2014)}, volume 1295 of {\em CEUR Workshop Proceedings},
	pages 46--50. CEUR-WS.org, 2014.
	
	\bibitem{aal_min_TKDE}
	{W.M.P. van der} Aalst, A.J.M.M. Weijters, and L.~Maruster.
	\newblock {Workflow Mining: Discovering Process Models from Event Logs}.
	\newblock {\em IEEE Transactions on Knowledge and Data Engineering},
	16(9):1128--1142, 2004.
	
	\bibitem{tonbeta166}
	{A.J.M.M.} Weijters, {W.M.P. van der} Aalst, and {A.K. Alves de} Medeiros.
	\newblock {Process Mining with the Heuristics Miner-algorithm}.
	\newblock BETA Working Paper Series, WP 166, Eindhoven University of
	Technology, Eindhoven, 2006.
	
	\bibitem{sander-tree-disc-PN2013}
	S.J.J. Leemans, D.~Fahland, and {W.M.P. van der} Aalst.
	\newblock {Discovering Block-structured Process Models from Event Logs: A
		Constructive Approach}.
	\newblock In J.M. Colom and J.~Desel, editors, {\em {Applications and Theory of
			Petri Nets 2013}}, volume 7927 of {\em Lecture Notes in Computer Science},
	pages 311--329. Springer-Verlag, Berlin, 2013.
	
	\bibitem{aal_sna_cscw}
	{W.M.P. van der} Aalst, H.A. Reijers, and M.~Song.
	\newblock {Discovering Social Networks from Event Logs}.
	\newblock {\em Computer Supported Cooperative work}, 14(6):549--593, 2005.
	
\end{thebibliography}

%\printindex{si}{Subject Index}
% \addtocontents{toc}{\protect\addvspace{3pc}}
%\cleardoublepage
%\printindex{oi}{Operator Index}

\end{document}